\documentclass[aps, prd, floats, floatfix,superscriptaddress, nofootinbib, twocolumn,showpacs]{revtex4}

\usepackage{graphicx}
\usepackage[usenames,dvipsnames]{color}
\usepackage[colorlinks, pdfborder={0 0 0}, plainpages=false]{hyperref}
\usepackage{amsmath, amssymb, amsfonts}
\usepackage{xspace} 
\usepackage{dcolumn}
\usepackage{bm}
\usepackage{multirow} 
\usepackage{ulem}
\usepackage{breakurl}

\newcommand{\fmin}{f_{\rm min}}
\newcommand{\fNyq}{f_{\rm Nyq}}
\newcommand{\MSun}{M_\odot}

\definecolor {darkgreen}{rgb}{0.2,0.7,0.2}

\definecolor{CiteColor}{rgb}{0, 0.5, 0}
\hypersetup{citecolor=CiteColor} %
\definecolor{RefColor}{rgb}{0.55, 0, 0}
\hypersetup{linkcolor=RefColor} %

\usepackage{ulem}
\begin{document}

\title{Systematic biases in parameter estimation of binary black-hole mergers}

\author{\surname {Tyson} B. Littenberg}\affiliation{Maryland Center
  for Fundamental Physics \& Joint Space-Science Institute, \\Department of Physics, University of
  Maryland, College Park, MD 20742}\affiliation{Gravitational
  Astrophysics Laboratory, NASA Goddard Spaceflight Center, \\8800
  Greenbelt Rd., Greenbelt, MD 20771}

\author{\surname {John} G. Baker}\affiliation{Gravitational
  Astrophysics Laboratory, NASA Goddard Spaceflight Center, \\8800
  Greenbelt Rd., Greenbelt, MD 20771}

\author{\surname {Alessandra} Buonanno}\affiliation{Maryland Center
  for Fundamental Physics \& Joint Space-Science Institute, \\Department of Physics, University of
  Maryland, College Park, MD 20742}

\author{\surname {Bernard} J. Kelly}\affiliation{Department of Physics, University of Maryland, Baltimore County, 1000 Hilltop Circle, Baltimore,
MD 21250}\affiliation{CRESST \& Gravitational Astrophysics Laboratory, NASA Goddard Spaceflight Center, 8800 Greenbelt Rd., Greenbelt, MD 20771}

\date{\today}

\begin{abstract}
  Parameter estimation of binary-black-hole merger events in
  gravitational-wave data relies on matched-filtering techniques,
  which, in turn, depend on accurate model waveforms.  Here we
  characterize the systematic biases introduced in measuring
  astrophysical parameters of binary black holes  by applying the currently most accurate
effective-one-body templates to simulated data containing non-spinning
  numerical-relativity waveforms.  For advanced ground-based
  detectors, we find that the systematic biases are well within the
  statistical error for realistic signal-to-noise ratio (SNR).  These biases grow to be comparable to the 
  statistical errors at high ground-based-instrument SNRs ($\rm{SNR}\sim50$), but never 
  dominate the error budget.
At the much larger signal-to-noise ratios expected for space-based detectors,
  these biases will become large compared to the statistical errors, 
  but for astrophysical black hole mass estimates the absolute biases (of at most a few percent) are still fairly small.
  \end{abstract}

\maketitle

\section{Introduction}
\label{intro}

Binary-black-hole (BBH) coalescences are cornerstone sources for gravitational-wave (GW) detectors, be they 
existing ground-based detectors like LIGO~\cite{LIGO} and Virgo~\cite{Virgo}, 
planned space-based detectors such as classic LISA~\cite{LISA} or eLISA~\cite{eLISA}, 
or a pulsar timing array~\cite{PTA}. The analysis of GW data to detect and characterize binary-black-hole merger events and to test the predictions of general relativity requires some family of efficiently computable signal models, representing all the possible waveforms consistent with general relativity.

Modeling BBH waveforms has historically been separated into three regimes, distinguished by the
different computational procedures suitable for each.  The
``inspiral'' where the individual black holes are sufficiently
separated for post-Newtonian (PN) theory to be valid~\cite{Sasaki:2003xr, Blanchet:2006zz, Futamase:2007zz,
  Goldberger:2004jt}, the
``merger'' of the binary where numerical relativity (NR) is
needed~\cite{Pretorius:2005gq, Campanelli:2005dd, Baker:2005vv}, and the 
``ringdown'' phase of a single, post-merger, perturbed object relaxing 
to a Kerr black hole~\cite{Vishveshwara:1970cc,Chandrasekhar:1975zza}. 

During recent years, work at the interface between analytical and numerical relativity 
has provided the community with a variety of semi-analytical inspiral-merger-ringdown waveform sets
~\cite{Buonanno:2006ui,Buonanno:2007pf,Pan:2007nw,Ajith:2007qp,Damour:2007yf,Damour:2009kr,
Ajith:2009bn, Pan:2009wj,Santamaria:2010yb, Sturani:2010yv,Pan:2011gk,Taracchini:2012ig} of varying
scope in parameter range and accuracy. These waveforms have already been used to search for GWs from high-mass~\cite{Abadie:2011kd, Aasi:2012dr}  
and intermediate-mass~\cite{Abadie:2012} binary black holes in LIGO and Virgo data, and also to carry out preliminary
parameter-estimation studies for ground-based detectors~\cite{Ajith:2009fz} and space-based detectors, such as 
classic LISA~\cite{McWilliams:2010eq, McWilliams:2011zs}. In this paper we shall study a set of inspiral-merger-ringdown waveforms based on the Effective-One-Body (EOB) framework~\cite{Buonanno:1998gg,
Buonanno:2000ef,Damour:2000we,Damour:2001tu,Buonanno:2005xu}. 

 Template waveforms will always be an approximation to the true 
 signals and the difference, if large enough, can bias inferences made from the GW data 
 about the astrophysical parameters of the system or the validity of general relativity.  
Estimates of the systematic errors introduced by waveform approximants in the 
literature \cite{Canitrot:2001hc,Cutler:2007mi,Bose:2008ix,Lindblom:2009ux,
Lindblom:2010mh,Pan:2011gk} have focused only on the inspiral, or used general 
conservative criteria to determine when the waveform has a bias, 
never using the parameter-estimation techniques employed in actual data analysis.

In this work, we will carry out the first \emph{measurement} of systematic
biases introduced when determining the physical parameters of a BBH merger 
by using EOB waveforms as templates.  We do so by simulating
data that contain NR waveforms as the ``signal'' to be detected. 
Then, using the Markov Chain Monte Carlo (MCMC) 
method~\cite{Metropolis:1953am,Hastings:1970}, we sample the posterior
distribution function for the binary parameters using EOB waveforms as
the templates.  The characteristic width of the posterior as determined
by the MCMC is taken as the \emph{statistical} error, while the
distance in parameter space between the dominant mode of the posterior and the
true, or ``injected'' waveform parameters is the \emph{systematic}
error, or bias, introduced by these waveforms. 
We study several different BBH systems, sampling the total mass,
mass-ratio, and signal-to-noise ratio (SNR) space for both a
LIGO/Virgo network in the advanced-detector era~\cite{aVirgo,aLIGO} and a LISA-like
configuration.

For this first analysis we employ non-spinning waveforms for 
quasi-circular orbits. In particular, for the injected signals, 
we consider the NR waveforms produced by the Caltech-Cornell-CITA collaboration 
in Ref.~\cite{Buchman:2012dw}. For the templates we use the EOB waveforms that were calibrated  
in Ref.~\cite{Pan:2011gk} to those NR waveforms~\cite{Buchman:2012dw}. 
Because our emphasis is on BBHs, and the merger
waveforms in particular, it is certainly the case that spin magnitude
and orientation play an important role in the waveform, and thus in parameter
estimation~\cite{Vecchio:2003tn,Berti:2004bd, Lang:2007ge, vanderSluys:2007st, Arun:2008zn}.
NR waveforms with spins aligned or anti-aligned with the orbital angular momentum are available, 
and EOB waveforms that include spins have been developed in 
Refs.~\cite{Buonanno:2005xu,Pan:2009wj,Taracchini:2012ig}. However, the spinning 
EOB waveforms are currently restricted to the dominant mode and 
additional code development is needed before they can be employed in stochastic sampling methods like the
MCMC. Thus, we leave to the future the extension of this study to spinning BBHs. 

Within these limitations, we show that the EOB waveforms developed in
Ref.~\cite{Pan:2011gk} and tested here are accurate enough to
introduce little to no significant biases when the data contain NR
waveforms at SNRs consistent with expectations for likely LIGO/Virgo
detections (${\rm SNR}\lesssim 50$).  For LISA-like detections, where
the expected SNRs are much higher than for ground-based detectors, statistically significant biases do emerge. 
Nonetheless, we 
find that the discrepancies between the true and measured parameters, 
at a few percent for the black-hole masses, are small enough to not impact key astrophysical
conclusions that may be drawn from the data (e.g., black-hole seed models,
etc.)~\cite{Plowman:2009rp,Sesana:2010wy, eLISA}.  However, when very high accuracies are required, as when 
testing the validity of general relativity~\cite{Li:2011cg,Vallisneri:2012qq}, best-fit EOB waveforms
from the existing model will leave behind significant residual power,
making them ill suited for these applications without further
development.

The remainder of the paper is organized as follows. 
In Sec.~\ref{sec:waveforms} we describe the numerical and analytic waveforms used 
in this work. In Sec.~\ref{sec:MCMC} we lay out how the study will proceed,
describing in particular the MCMC sampler that we 
use. We then discuss in detail the results for stellar-mass BBHs in ground-based
detectors (Sec.~\ref{sec:LIGO}) and supermassive BBHs in space-based
observatories (Sec.~\ref{sec:LISA}).  In Sec.~\ref{sec:conclusions} we
summarize the findings from this work, address limitations, and
discuss future directions to be pursued.

\section{Inspiral-merger-ringdown  waveforms used in the analysis}
\label{sec:waveforms}

Our study involves comparisons between two sets of waveforms.  We primarily seek to
\emph{evaluate} a continuously parameterizable family of model waveforms based on
the EOB framework against a discrete set of highly accurate numerical relativity (NR)
waveforms. In analogy with observational algorithm tests, we can think of the numerical waveforms as ``injected'' signals,
which we challenge the ``template'' EOB waveforms to match.

We employ as injected signals the non-spinning NR waveforms 
produced by the Caltech-Cornell-CITA collaboration~\cite{Buchman:2012dw},  
using the spectral Einstein code. The NR polarizations have mass-ratio $q \equiv m_1/m_2 = 1, 2, 3, 4, 6$ and 
contain -2 spin-weighted spherical harmonics $(\ell,m) = (2,\pm 2)$,
$(2,\pm1)$, $(2,0)$, $(3,\pm 3)$, $(3,\pm 2)$, $(4,\pm4)$, $(5,\pm5)$, and $(6,\pm6)$.
These waveforms provide $30\mbox{--}40$ (quadrupole) GW cycles before merger, 
depending on the mass-ratio. 

The phase and amplitude errors of the NR waveforms
vary with mass-ratio and gravitational mode. The numerical errors grow toward merger and ringdown, and  
typically at merger, for the dominant $(2,2)$ mode, the phase error ranges between 
0.05 and 0.25 rad, while the fractional amplitude error is at most $1\%$. The subdominant 
modes can have somewhat larger errors, especially the $(3,3)$ and $(4,4)$ modes. 

Applying the numerical waveforms to generate mock signal observations, we will test the ability of
a previously published family of template waveforms to characterize these signals \cite{Pan:2011gk}. 
These template waveforms are based on the EOB framework, founded on the very accurate results of 
PN theory, an expansion of general-relativity dynamics in powers of $v/c$, where 
$v$ is the characteristic velocity of the binary. In the EOB approach, however,
the PN expansions are applied in a resummed form that maps the dynamics of two compact objects
into the dynamics of a reduced-mass test particle moving in a deformed Schwarzschild 
geometry~\cite{Buonanno:1998gg,Buonanno:2000ef,Damour:2000we,Damour:2001tu,Buonanno:2005xu}. Waveforms
in the EOB formalism are derived from such particle dynamics up to the light-ring (unstable photon
orbit) radius. The subsequent ringdown portion of the waveforms is a superposition of quasinormal
modes matched continuously to the inspiral. Tunable parameters, effectively standing in for currently 
unknown higher-order PN terms, are fixed by matching to numerical-relativity simulation results.

The comparable-mass NR waveforms in Ref.~\cite{Buchman:2012dw} were
used, together with the small-mass-ratio waveforms produced by the
Teukolsky code in Ref.~\cite{Barausse:2011kb}, to calibrate a
non-spinning EOB model in Ref.~\cite{Pan:2011gk}. More specifically,
the numerical waveforms available at discrete points in the parameter
space were employed to fix a handful of EOB adjustable parameters
entering the EOB conservative dynamics and gravitational modes.  These
adjustable parameters were then interpolated over the entire
mass-ratio space.  The EOB model in  Ref.~\cite{Pan:2011gk} contains four subdominant
gravitational modes, $(2,\pm 1), (3,\pm 3), (4,\pm 4)$ and $(5,\pm
5)$, beyond the dominant mode $(2,\pm 2)$.

The EOB model in Ref.~\cite{Pan:2011gk} has been coded in the (public) LIGO Algorithm Library (LAL)~\cite{LAL} 
(under the name EOBNRv2). We carry out our study using LAL to generate template waveforms. Henceforth, we denote the EOB model with only 
the dominant $(2,\pm 2)$ mode as EOB$_{22}$, and the model that includes the four subdominant 
modes $(2,\pm 1), (3,\pm 3), (4,\pm 4)$ and $(5,\pm 5)$ as EOB$_{\rm HH}$.
We will also omit the $\pm$ in mode labels.

The phase difference of the $(2,2)$ mode between the calibrated EOB model and numerical simulation 
remains below $\sim 0.1$ rad throughout the evolution for all mass-ratios considered; 
the fractional amplitude difference at merger (i.e., at the waveform's peak) of 
the $(2,2)$ mode is $2\%$, growing 
to $12\%$ during the ringdown. Around merger and ringdown, the phase and amplitude differences of the subdominant modes 
between the EOB and NR waveforms are somewhat larger than those of the $(2,2)$ mode. 
[The numerical errors, and phase and amplitude differences between the EOB and NR 
waveforms can be read off from Figs. 6--10 in Ref.~\cite{Pan:2011gk}.]

To quantify how these differences between template and signal would affect GW searches in Advanced 
LIGO, Ref.~\cite{Pan:2011gk} studied the effectualness and measurement accuracy of the EOB model. 
When investigating the effectualness for detection purposes, they 
found that the NR polarizations containing the
strongest seven modes~\footnote{Note that in Ref.~\cite{Pan:2011gk} the $(2,0)$ mode was 
not included.} have a maximum mismatch of $7\%$ for stellar-mass 
BBHs, and $10\%$ for intermediate-mass BBHs, when 
using only EOB$_{22}$ for $q = 1, 2, 3, 4, 6$ and
binary total masses $20\text{--}200$ $\MSun$.  However, the mismatches
decrease when
using the full EOB$_{\rm HH}$ model, reaching an upper bound of $0.5\%$ for
stellar-mass BBHs, and $0.8\%$ for intermediate-mass BBHs. 
Thus, the EOB model developed in Ref.~\cite{Pan:2011gk} is accurate enough 
for detection, which generally requires a mismatch not larger than $7\%$. 

To understand whether this EOB model is precise
enough for measurement purposes, the authors of Ref.~\cite{Pan:2011gk} carried out a preliminary
study, adopting as accuracy requirement for measurement the one
proposed in Refs.~\cite{Lindblom:2009ux,Lindblom:2010mh}. Using a
single Advanced LIGO detector, Ref.~\cite{Pan:2011gk} computed the SNRs below which
the EOB polarizations are accurate enough that systematic biases are
smaller than statistical errors. Since subdominant modes have
non-negligible contribution for large mass-ratios, and those modes
have the largest amplitude errors, they found that the upper-bound
SNRs are lower for the most asymmetric systems, such as $q=6$.
However, as stressed in Ref.~\cite{Pan:2011gk}, the accuracy requirement in
Ref.~\cite{Lindblom:2009ux,Lindblom:2010mh} may be too conservative,
and by itself does not say which of the binary parameters 
will be biased and how large the bias will be. It could turn
out that the biased parameters have little relevance in astrophysics
or tests of general relativity.  It is the main goal of this paper to 
measure the actual biases of the EOB model with and without the
subdominant modes.

Our study is restricted to binary systems moving along quasi-circular orbits where
the spin of each constituent black hole is negligible, thus reducing
the model parameters $\boldsymbol\theta$ from a space of 17 dimensions 
to 9 dimensions:
\begin{equation}
\label{par}
\boldsymbol\theta = 
\left\{\ln M, \ln \mathcal{M},  \ln D_{\rm L}, t_{\rm p}, \sin \delta, \alpha, \cos \iota, \psi, \varphi_{\rm p} \right\}\,.
\end{equation}
In the above parameter list, $M \equiv (1+z) (m_1 + m_2)$ is the redshifted total mass
of the binary, and $\mathcal{M} \equiv \nu^{3/5}\, M$ is its chirp mass,
where $\nu \equiv m_1 m_2/(m_1 + m_2)^2 = q/(1+q)^2$ is the symmetric mass-ratio.
We
denote by $D_L$ the luminosity distance, which, along with the right
ascension $\alpha$ and declination $\delta$, describes the location of
the binary. The orientation of the binary's orbital
angular-momentum vector $\hat{\mathbf{L}}$ with respect to the line of sight
$\hat{\mathbf{k}}$ from the observer is encoded in the model using the 
inclination $\iota$, polarization angle $\psi$, and phase
$\varphi_{\rm p}$ --- the Euler angles that describe the rotation
from $\hat{\mathbf{k}}$ to $\hat{\mathbf{L}}$. The parameter $t_{\rm p}$ 
is the time of the $(2,2)$ mode's maximum amplitude,
a proxy for the binary merger time, and $\varphi_{\rm p}$ is the GW phase at $t_{\rm p}$.

Our comparisons between the EOB waveforms and the NR data are restricted to the late-inspiral, merger and 
ringdown, that is roughly $30\mbox{--}40$ GW cycles before 
merger, depending on the mass-ratio. The injected signals contain only 
the NR waveforms available. We do not match the NR waveforms to EOB or PN waveforms 
at low frequency to increase the number of cycles, because we do not know how 
well the EOB or PN waveforms would approximate the NR waveforms outside the 
region of calibration and we do not want to introduce unknown errors 
when estimating the systematic biases. 

There is no guarantee that a template that is in phase with the NR waveform during
the last $30\mbox{--}40$ GW cycles will remain so throughout the entire
inspiral. The mass parameters are strongly encoded in the GW phase, 
so any additional de-phasing at times earlier than 
covered by the NR simulations can potentially increase the systematic
biases.  Therefore, in order for our study to be meaningful, 
we consider binary systems with a total mass $M$ such that the majority of 
the SNR is accumulated during  the last $30\mbox{--}40$ GW cycles before merger. 

\section{Statistical versus systematic errors using MCMC techniques}
\label{sec:MCMC}

We use the MCMC algorithm~\cite{Metropolis:1953am,Hastings:1970} to produce samples from the posterior
distribution function for the waveform parameters.  The MCMC sampler is built on the foundation of Bayes' Theorem, 
which, in the context of parameter inference, defines the posterior distribution function for parameter vector $\boldsymbol\theta$ and data $d$ as
\begin{equation}\label{eq:bayes}
p(\boldsymbol\theta|d,\mathcal{I}) \equiv \frac{p(d|\boldsymbol\theta,\mathcal{I})p(\boldsymbol\theta|\mathcal{I})}{p(d|\mathcal{I})}.
\end{equation}
Here $p(\cdot | \cdot)$ are conditional probability densities with arguments on the right-hand side 
of the bar assumed to be true, $p(d|\boldsymbol\theta,\mathcal{I})$ is the likelihood, 
$p(\boldsymbol\theta|\mathcal{I})$ is the prior distribution, and $p(d|\mathcal{I})$ is the model evidence, which, in 
parameter-estimation applications, serves only as a normalization constant.  The information $\mathcal{I}$ denotes 
all of the assumptions that are built into the analysis, particularly that the NR waveforms represent 
reality (see Sec.~\ref{sec:waveforms} for associated discussion).
Henceforth, to simplify notation, we shall not include $\mathcal{I}$ when writing the conditional 
probability density $p(\cdot|\cdot)$. When comparing different model combinations, we adopt the 
notation for conditional probabilities with arguments to the right of a vertical bar representing the data and
arguments to the left signifying which model was used as templates.
For example, results labeled $p(\rm{EOB}_{22}|\rm{NR})$ come from the
posterior distribution functions for models using $\rm{EOB}_{22}$ as
the templates and an NR waveform as the data.

With the posteriors, we compare the statistical 
error (the characteristic width of the posterior) to the systematic error (the displacement from the
injected parameter values of the posterior's mode).  We do not include
a noise realization in the simulated data, as that introduces
additional biases --- each noise realization pushing the best-fit
solution away from the injected value in a different way --- that are
not easily quantified~\cite{Raymond:2009cv}.  As the control in this
experiment, we simulated (noise-free) data with EOB waveforms and use
EOB templates for parameter estimation, giving us one set of results
with no systematic bias, apart from the sampling error in the Markov
chains due to their finite length.  We use these controlled results as
code verification, and as the standard against which the other models'
performance is compared.  We then test the EOB waveform models by
injecting NR waveforms (summed over all available modes) and using the
two EOB models discussed in Sec.~\ref{sec:waveforms} as templates.
The EOB$_{22}$ model is used as a baseline, as it
has been employed in LIGO/Virgo search pipelines to analyze data
collected in recent science runs~\cite{Abadie:2010yb, Abadie:2012}. 
The EOB$_{\rm  HH}$ model is the most complete waveform
at our disposal, and is used to measure how well the EOB model could
perform on NR data.

In Eq.~(\ref{eq:bayes}) we use the standard gaussian logarithmic likelihood
 $\ln p(d|\boldsymbol\theta)
\equiv -(d-h(\boldsymbol\theta)|d-h(\boldsymbol\theta))/2 + C$,
where $C$ is a normalization constant that does not depend on model 
parameters and is henceforth neglected, $h$ is the template, and
\begin{equation}
(a|b)\equiv 2\sum_{i=1}^{I}\int_{\fmin}^{\fNyq} 
\frac{\tilde{a}^*_i(f)\tilde{b}_i(f) + \tilde{a}_i(f)\tilde{b}^*_i(f)}{S_n^i(f)}df\,,
\label{eq:nwip}
\end{equation}
denotes a noise-weighted inner product with the sum on $i$ over $I$
(independent) interferometer channels and $S_n^i(f)$ is the one-sided
noise power spectral density (PSD) for detector $i$. The bounds of
integration, $\fmin$ \& $\fNyq$, are the minimum frequency of
the NR waveform and the Nyquist frequency of the data, respectively.
The Nyquist frequency is chosen to ensure that the highest-frequency
portion of the waveform is well below the instrument sensitivity
curve, while $\fmin$ is set by the duration of the NR waveform
and the total mass of the system, such that we only integrate over
frequencies where numerical data exists.

For each case, the Markov chains are run for $\sim 10^6$ iterations, taking about
$10^3$ CPU hours to complete. The chains rely on parallel
tempering~\cite{Swendsen:1986}, differential
evolution~\cite{DiffEvol}, and jumps along eigenvectors of the Fisher Information Matrix (FIM) 
(e.g., see Ref.~\cite{Cornish:2005qw}) computed from PN
waveforms to efficiently explore the posterior distribution function.
We use burn-in times of $10^4$ samples, and run several chains with
different initial locations to check for convergence.  Prior
distributions for all parameters are chosen to be uniform.  Azimuthal angular
parameters ($\alpha$, $\psi$, and $\varphi_p$) have support over $[0,2\pi)$ with periodic boundary conditions,
while declination-like angle parameters ($\sin\delta$, $\cos\iota$) range from $[-1, 1]$ with reflecting boundary conditions. The ranges for
$\ln M$ and $\ln D_{\rm L}$ are chosen to be large enough so as to not influence
the posteriors.  The prior range on $\ln \mathcal{M}$ is coupled to
$\ln M$, as the maximum value of the chirp mass occurs for the $q=1$
($\nu=1/4$) case, and depends on the total mass $M$ of the system.  Because of this,
the prior boundary on chirp mass does affect the posteriors for the
equal-mass systems considered in this work.
  
The products of our analysis procedure are samples from the posterior
distribution function $p(\boldsymbol\theta|d)$ --- an oddly shaped, sometimes multimodal, blob
living in a 9D space.  There is no perfect way of
distilling this information into a simple, robust, statistic to assess
parameter-estimation accuracy.  We will make do with the ``fractional
systematic error" $\delta\beta_{\theta}$:  For parameter $\theta$, we
first define the systematic error $\beta_\theta \equiv |\theta_{\rm MAP} - \theta_0|$ 
where ${\theta}_{\rm MAP}$ is the \emph{maximum a posteriori} (MAP) value
and $\theta_0$ is the injected value, while the
statistical error is quantified by the standard deviation $\sigma_\theta$ of the 1D 
marginalized posterior distribution function.  We
then define the fractional systematic error as the ratio between
$\beta_{\theta}$ and the statistical error:
\begin{equation}\label{eq:delta_beta}
\delta\beta_{\theta} \equiv \frac{\beta_\theta}{\sigma_{\theta}}\,.
\end{equation}
We consider templates that consistently yield $\delta\beta_\theta
\lesssim 1$ as introducing negligible bias, assuming the NR waveforms 
are exact, which, as seen in Sec.~\ref{sec:waveforms}, is not the case~\footnote{  
Currently we have no way of incorporating the numerical error 
of the NR waveforms into our estimate and so we neglect it.}. 

The fractional systematic error \eqref{eq:delta_beta} can be interpreted as the number of
standard deviations away from the injected value at which we find the
MAP waveform. This choice of statistic is not perfect --- low-SNR
systems have very non-gaussian posteriors making the standard
deviation a poor choice for characterizing the statistical error.
Furthermore, the MAP parameters are a single point and tell us nothing
about how large a region in parameter space had similar posterior
support to the current best estimate.  Additionally, the MAP value is
a feature of the full 9D posterior, while the variances are computed
from the \emph{marginalized} posterior distribution functions.  This
introduces complications for some special cases, as we shall discuss in 
detail below.

\section{Results for Advanced LIGO detectors}
\label{sec:LIGO}

The first test of the EOB waveforms uses simulated data from the
network of advanced ground-based detectors expected to come on line in
the middle of this decade: the two LIGO detectors in the USA and the
Virgo detector in Italy.  We use the same noise PSD for each
interferometer, the ``zero-detuned high-power'' curve from
Ref.~\cite{PSD}, which is the sensitivity curve for the fully completed Advanced LIGO
detector.  The GW response in each interferometer
is modeled by convolving the GW signal with the
beam-pattern function for that detector and applying the appropriate 
time-delays between interferometers~\cite{Anderson:2001}.

\subsection{Choice of binary configurations}

We study several binary configurations using different mass-ratios, total masses,
and SNRs.  The SNR of the system is computed via
${\rm SNR} = \sqrt{(d|d)}$, and its value is controlled by adjusting
the luminosity distance $D_L$.  Because our simulated data $d$
contain no simulated noise, the SNR is simply the inner product of
the injected waveform with itself.  Also, our definition of the inner
product in Eq.~(\ref{eq:nwip}) includes a summation over all
interferometer channels, thus we quote the  \emph{network} SNRs for the
ground-based studies.

\begin{figure*}[htbp]
   \centering
      \includegraphics[width = 1\linewidth]{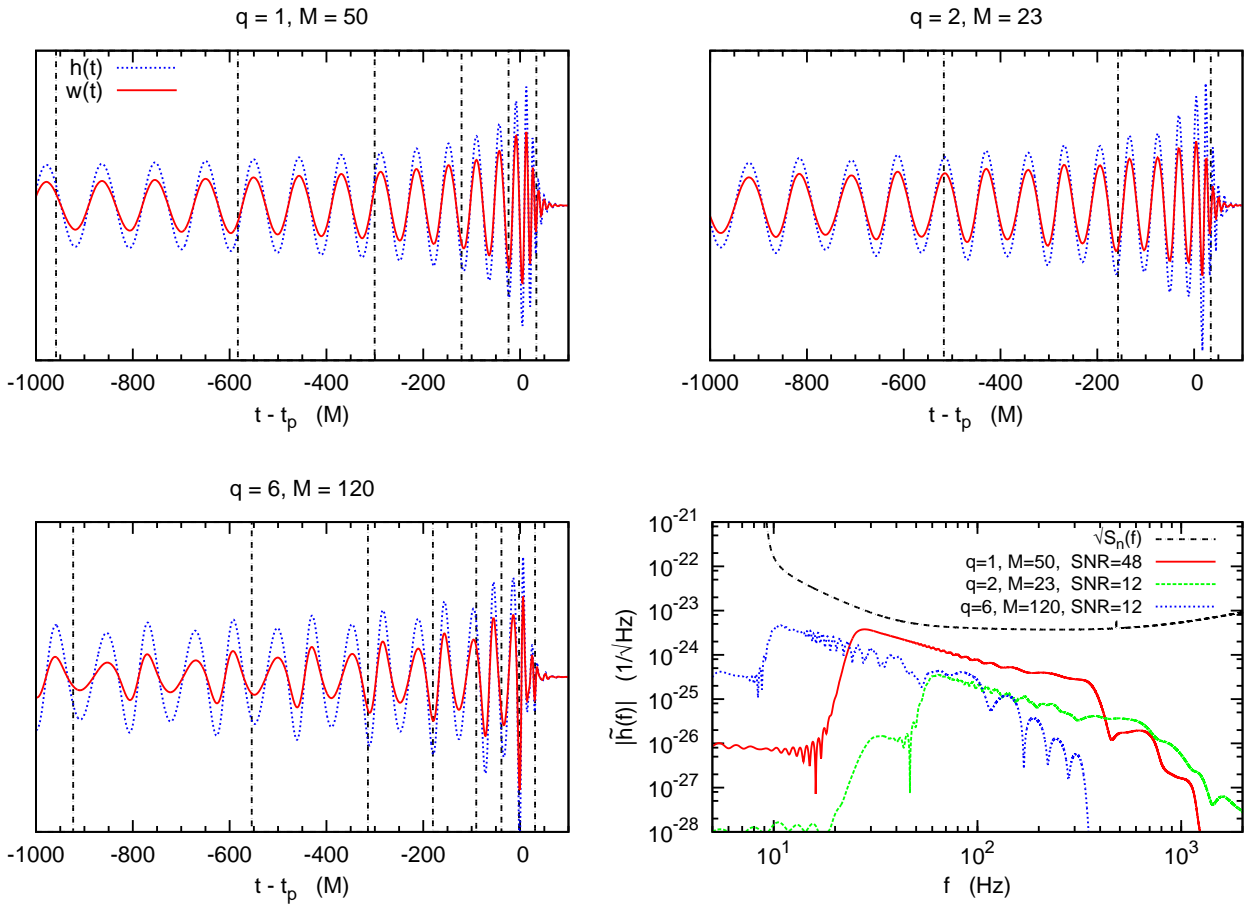} 
   \caption{{\small 
       [Top row, and bottom-left] Time-domain EOB waveforms (blue dotted) and 
       the same signal whitened by the noise spectral density (red solid) representing different test cases studied 
       in this work.  The time-axis is scaled by the total mass $M$ and shifted by the 
       merger time $t_p$.  The horizontal dotted lines demarcate $10\%$ intervals for 
       accumulated signal power, with the rightmost line at the $99.9\%$ mark.}
       [Bottom right] Strain spectral densities $|\tilde{h}(f)|$ showing
       the estimate for the Advanced LIGO noise curve (gray dotted
       line) and the NR waveforms for the same representative examples as the other panels.
       }
   \label{fig:ligo_psd}
\end{figure*}
The first three panels of Fig.~\ref{fig:ligo_psd} show the time-domain 
EOB$_{\rm HH}$ waveforms (blue, dotted) and whitened by the noise spectral density (red, solid) 
for three representative cases studied here\footnote{``Whitened'' waveforms are ones that have been
Fourier-transformed to the frequency domain, rescaled by $1/\sqrt{S_n(f)}$, and finally
re-transformed to the time domain~\cite{Damour:2000gg}.}.  The vertical lines indicate intervals in 
which contribute $10\%$ of the signal power, starting from $\fmin$ for each system, with the rightmost line indicating where 
$99.9\%$ of the power has accumulated.
We focus on the late-inspiral, merger, and ringdown portions of the waveform, beginning 1000 $M$ before 
the peak of the $(2,2)$ mode's amplitude.
The power intervals are included as a guide to see which portions of the waveform contribute most 
to the parameter estimation.  For instance, 50\% of the $q=1,\ M=50\ \MSun$ waveform's integrated power is 
contained in the late-inspiral, merger, and ringdown, while the $q=2,\ M=23\ \MSun$ signal 
is much less dominated by this interval, accounting for only 30\% of the power.
The $q=6,\ M=120\ \MSun$ examples are most influenced by the end of the waveform, 
which makes up over 70\% of the power.

The bottom-right panel of Fig.~\ref{fig:ligo_psd} shows the strain spectral densities $|\tilde{h}(f)|$ 
for the same three systems, now using the NR waveforms used in this study. 
Also included is the Advanced LIGO power spectral density.  

We focus on moderately high-mass black-hole mergers with $M\sim50\ \MSun$ (e.g., the equal-mass case 
shown in Fig.~\ref{fig:ligo_psd}: top-left panel, red
solid curve in bottom-right panel). Beyond their potential as Advanced LIGO sources\footnote{ 
A binary with $M\sim\ 50\MSun$ is astrophysically relevant, as the  
largest black-hole 
mass ever observed is in the range of $23\mbox{--}34\ \MSun$~\cite{Prestwich:2007mj,Silverman:2008ss}. 
Recent results from population-synthesis studies suggest that massive, low-metallicity stars 
are capable of producing black holes as large as $M\sim80\ \MSun$~\cite{Belczynski:2009xy}, although these 
findings are for single stars only, and binary evolution could either increase or decrease the maximum black-hole 
mass. Additionally, there exists at least one example of a massive Wolf-Rayet star, 
R136a1~\cite{Crowther:2010cg}, with $M\sim\ 250\MSun$ at a distance of $\sim0.1$ Mpc and with sufficiently low
metallicity to produce a massive black hole. However, it cannot be excluded that the star goes instead 
through a pair-instability supernova, leaving no remnant.}, high-mass systems serve an important role in testing the waveform models for two
reasons. First, as explained in Sec.~\ref{sec:waveforms}, the NR signals are short 
in duration and we do not supplement the waveform by hybridizing the numerical 
data with analytic inspiral models at low frequency. We therefore require higher-mass systems, 
merging at lower frequency, to ensure that most of the
inspiral missing from the NR data will fall outside the sensitive measurement
band of the detector.  
With $M\sim50\ \MSun$, NR waveforms start at $\sim30$ Hz, setting $\fmin$ in the inner product defined in 
Eq.~(\ref{eq:nwip}).  Comparing these data to EOB waveforms with the 
same parameters but $\fmin=10$ Hz (below which the Advanced LIGO sensitivity 
is very poor), we find the NR waveforms contain $\sim85\%$ of the total 
signal power, or $\sim90\%$ of the total SNR.

A second reason for focusing on $M\sim50\ \MSun$ for the 
binary is that these systems are
``centrally located'' in frequency over the most sensitive band of the
advanced detectors ($\sim 30$ to $\sim 10^3$ Hz, e.g., see the bottom-right panel in Fig.~\ref{fig:ligo_psd}), 
such that inspiral, merger, ringdown, and additional modes all contribute 
to the overall signal power and, accordingly, the parameter-estimation capabilities. 
Generally speaking, we expect such systems to make the greatest demands on complete 
inspiral-merger-ringdown waveform model accuracy.

While the $M\sim50\ \MSun$ systems serve as the basis for our
comparisons, we include additional examples to probe regions of
signal space that are of particular interest.  These include a $q=6$,
$120\MSun$ system 
(Fig.~\ref{fig:ligo_psd}:  bottom-left panel, blue dotted curve in bottom-right panel) 
chosen such that the subdominant 
modes contribute the most, as they will be most
pronounced at high mass-ratios, and signal power from the higher
frequency modes is still in the sensitive band of the detector.  
At this mass, $\fmin=10$ Hz, so our analysis is not missing any 
signal power due to the length of the NR data.

We also go to lower masses, using a $q=2$, $M=23\ \MSun$ binary 
(Fig.~\ref{fig:ligo_psd}:  top-right panel, green dashed curve in bottom-right panel) 
as a more likely LIGO/Virgo detection, to demonstrate
the EOB models' parameter-estimation accuracy not at the extremes of a
potential binary signal, but within reasonable expectations of what
the coming data may hold (apart from including the black-hole spins). 
It is worth noting that for these low-mass systems, we are missing a large 
portion of the inspiral, as $\fmin=60$ Hz and 
$\sim30\%$ of the full SNR will be accumulated below that frequency.  Therefore, these results might change 
in the future, when longer EOB and NR waveforms become available.

\subsection{Results on systematic biases at fixed inclination angle}

\begin{figure*}[htbp]
   \centering
   \includegraphics[width = 1\textwidth]{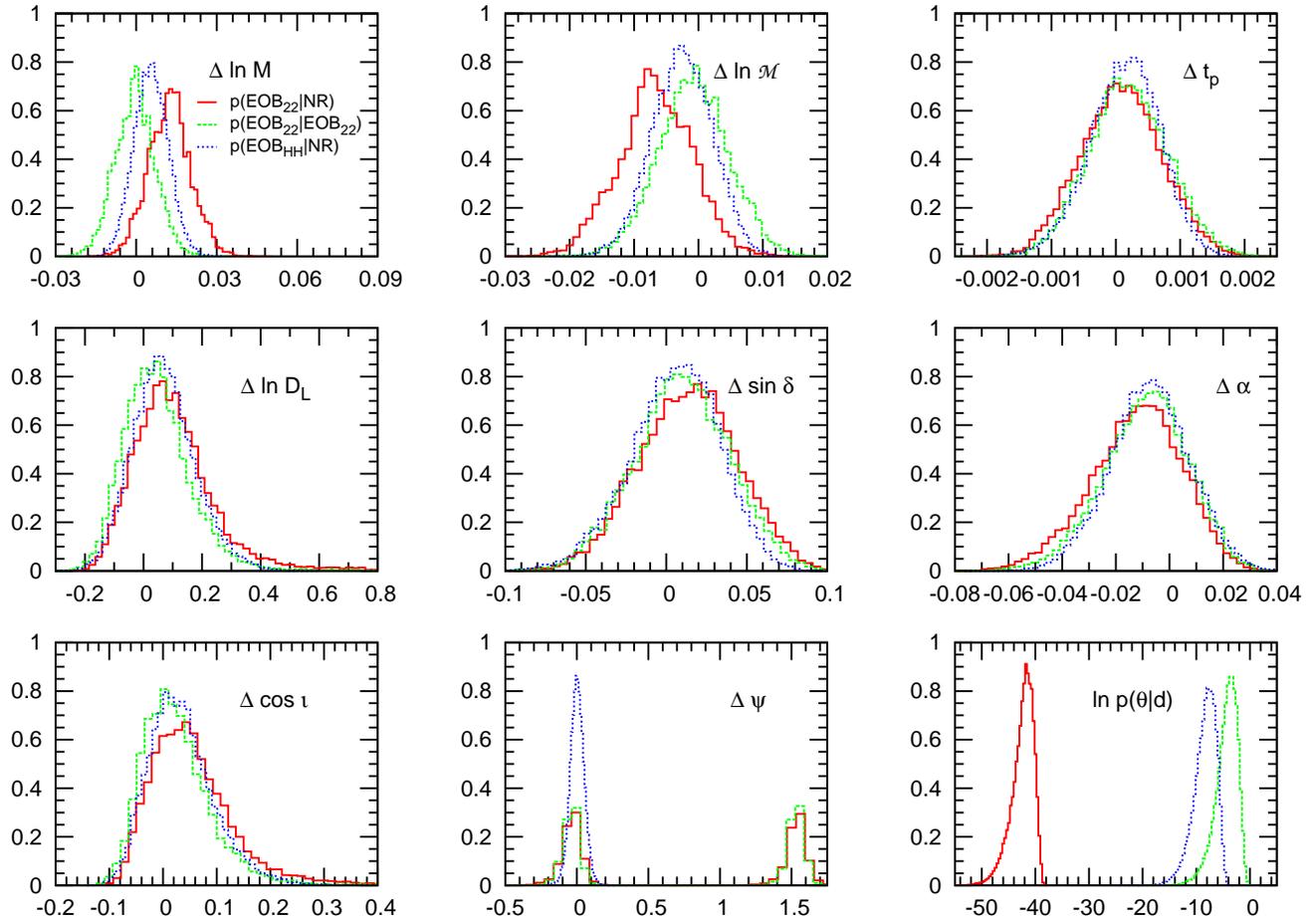} 
   \caption{{\small Example marginalized posterior distribution
       functions for a binary system with mass-ratio $q=2$, total mass
       $M = 51\MSun$, and network ${\rm SNR} = 48$ produced using an MCMC sampler.  
       The green (dashed)  histograms arise from injecting the EOB$_{22}$ waveforms 
       and using the same model as the template.  The red (solid)
       distributions use the EOB$_{22}$ waveform  while the blue
       (dotted) curves use the EOB$_{\rm HH}$ waveform to recover an NR injection.  The $x$-axis
       shows the distance away from the injected parameter value, so
       distributions that peak at zero show no bias.  
       The bottom-right  panel shows the logarithmic likelihood distributions, which are used to
       quantify the amount of residual power left behind by the
       waveform model.  The $y$-axes have be normalized to range between $[0,1]$.}}
   \label{fig:posteriors}
\end{figure*}
In Fig.~\ref{fig:posteriors} we plot the 1D marginalized posterior distribution 
functions for each parameter for the case of a binary 
with mass-ratio $q=2$, total mass $M = 51\ \MSun$, and network ${\rm SNR} 
= 48$, produced using an MCMC sampler. The inclination angle is chosen to be 
$\iota = \pi/3$. 
The independent variables in these plots are
$\boldsymbol\Delta \boldsymbol\theta = \boldsymbol\theta - \boldsymbol\theta_{\rm inj}$, where
$\boldsymbol\theta_{\rm inj}$ are the injected parameter values.
The $p({\rm EOB_{22}}|{\rm EOB_{22}})$ histograms (green, dashed lines) are the posteriors using the 
EOB$_{22}$ waveforms for both the signal and the templates. These confirm that the MCMC sampler is 
working properly, as the posteriors all show strong support for the 
injected waveform parameters (peaking at or near $0$), and statistical errors 
consistent with results in Ref.~\cite{Ajith:2009fz} obtained using FIM estimates and 
phenomenological inspiral-merger-ringdown waveforms.

The red (solid) lines and blue (dotted) lines are for data containing an NR
signal and the EOB$_{22}$ and EOB$_{\rm HH}$ waveforms as templates,
respectively.  The bottom-right panel shows the logarithmic likelihood
distributions for each chain.  We can see from these posteriors that
the EOB$_{22}$ waveform is significantly biased away from the NR 
injected value, by $\sim1\%$ in both $M$ and $\mathcal{M}$.  This bias
is substantially reduced when using the EOB$_{\rm HH}$ template,
to the point where the systematic error is well
within the statistical error of the posterior.  For the extrinsic
parameters such as distance and sky location, the posteriors for the
approximate templates are nearly identical to those produced by using
the exact same waveform for both data simulation and parameter
estimation.  We also see the role that subdominant modes play in
breaking the $\pi/2$ degeneracy in the polarization angle $\psi$, which
can aid in distance and sky-location determination for some systems.

The SNR of the residual $d-h$, given by ${\rm SNR}_{\rm res} = \sqrt{-2\ln 
p(d|\boldsymbol\theta)}$ can be inferred from the bottom-right panel. 
Viewed in this way, there is a distinct excess in the residual for even the
EOB$_{\rm HH}$ case (blue, dotted) in comparison with the idealized
control MCMC residuals (green, dashed). To understand the significance of this,
consider applying a detection threshold of $\rm{SNR}= 6$~\cite{Cutler:1992tc} for the residual waveform.
This corresponds to a maximum logarithmic likelihood of $\lesssim-18$, below which the residual could
potentially contain enough power to be detected after the best-fit waveform is regressed from the data.  
Suppose in the near future we have model waveforms at our disposal containing all of the details of black-hole  
mergers (i.e. spins and eccentricity) that generally produce $\rm{SNR}_{\rm res} \lesssim 6$ 
for equally detailed NR simulations, and yet coherent residuals are consistently found in the data. 
Such an event could suggest a possible departure from general relativity. 

\begin{table*}[htdp]
\begin{center}
\begin{tabular}{lccc|cccc|cccc|cccc}
\hline\hline
&&&& \multicolumn{4}{c|}{$p(\rm{EOB}|\rm{EOB})$} &  \multicolumn{4}{c|}{$p(\rm{EOB_{22}}|\rm{NR})$} & \multicolumn{4}{c}{$p(\rm{EOB_{HH}}|\rm{NR})$} \\ 

$q$ & $M$ ($\MSun$) & $f_{\rm low}$ (Hz)& SNR & $\sigma_{\ln M}$ & $\sigma_{\ln \mathcal{M}}$ & $\delta\beta_{\ln M}$ & $\delta\beta_{\ln \mathcal{M}}$ & $\sigma_{\ln M}$ & $\sigma_{\ln \mathcal{M}}$ & $\delta\beta_{\ln M}$ & $\delta\beta_{\ln \mathcal{M}}$ & $\sigma_{\ln M}$ & $\sigma_{\ln \mathcal{M}}$ & $\delta\beta_{\ln M}$ & $\delta\beta_{\ln \mathcal{M}}$ \\
\hline
1 & 50& 30 & 12 & 0.02 & 0.02 & 0.11 & 0.05 & 0.02 & 0.02 & 0.34 & 0.17 & 0.02 & 0.02 & 0.29 & 0.10\\

1 & 50& 30 & 48 & $4\times10^{-3}$ & $3\times10^{-3}$ & $1\times10^{-3}$ & 0.14 & 0.01 & 0.01 & 1.76 & 0.84 & $2\times10^{-3}$ & $2\times10^{-3}$ & 0.79 & 0.87\\

2$^*$ & 23& 60 & 12 & 0.02 & 0.01 & 0.01 & 0.02 & 0.03 & 0.02 & 0.27 & 0.07 & 0.03 & 0.02 & 0.18 & 0.26\\

2 & 51& 30 & 12 & 0.03 & 0.02 & $3\times10^{-3}$ & $4\times10^{-3}$ & 0.03 & 0.02 & 0.47 & 0.28 & 0.03 & 0.02 & 0.01 & 0.01\\

2 & 51& 30 & 48 & 0.01 & 0.01 & 0.01 & 0.01 & 0.01 & 0.01 & 1.92 & 1.39 & 0.01 & 0.01 & 0.93 & 0.32\\

6 & 56& 30 & 12 & 0.03 & 0.03 & 0.03 & 0.08 & 0.03 & 0.03 & 0.94 & 0.66 & 0.02 & 0.02 & 0.58 & 0.39\\

6 & 56& 30 & 48 & 0.01 & $5\times10^{-3}$ & 0.23 & 0.04 & 0.01 & 0.01 & 3.47 & 2.51 & 0.01 & $4\times10^{-3}$ & 1.43 & 0.84\\

6$^*$ & 120& 10  & 12 & 0.03 & 0.03 & 0.05 & 0.14 & 0.03 & 0.03 & 1.67 & 0.60 & 0.02 & 0.02 & 0.29 & 0.06\\
\hline
\end{tabular}
\caption{{\small Fractional systematic biases $\delta\beta$
    [see Eq.~(\ref{eq:delta_beta})] and statistical errors $\sigma$ for
    intrinsic parameters as determined by the MCMC sampler. 
    An asterisk in the mass-ratio column indicates examples where 
     EOB$_{\rm HH}$ was used for the  $p({\rm EOB}|{\rm EOB})$ study.
    All other examples used the EOB$_{22}$ waveform. }}
\label{tbl:intrinsic}
\end{center}
\end{table*}
The results from our MCMC studies for different systems are displayed
in Tables~\ref{tbl:intrinsic} and \ref{tbl:extrinsic}, which show
the fractional systematic error $\delta\beta$ (see Eq.~\eqref{eq:delta_beta})
for the mass, sky location, and distance parameters.
The injected waveforms again had an inclination angle of $\iota = \pi/3$.
We include our estimate of the statistical error $\sigma_\theta$ to
quantify the precision of the advanced detectors for the systems
considered here.  The standard deviations should be interpreted with
caution; we do not include noise in the simulated data, so the deviations are
not representative of the ``error bars'' on a \emph{particular}
detection, but instead represent an ensemble average over idealized
Gaussian stationary noise realizations of the
statistical error for these particular systems.

Table \ref{tbl:intrinsic} contains the \emph{intrinsic} parameters ---
those that affect the shape of the waveform.  Because we consider
non-spinning black holes, the only intrinsic parameters are the
masses.  The \emph{extrinsic}, or observer-dependent, parameters
(i.e., distance and sky location) are given in Table~\ref{tbl:extrinsic}. 
They are encoded in the instrument response to the GW, instead of being
imprinted in the phase and amplitude evolution of the waveform itself.  
We do not report on the orientation parameters $\iota$ and $\psi$ 
or reference time $t_{\rm p}$ and phase $\varphi_{\rm p}$ parameters in 
this fashion, but note that results for these other parameters 
are consistent with the extrinsic variables in Table~\ref{tbl:extrinsic}.

From Table~\ref{tbl:intrinsic} we can see that generically, the
EOB$_{22}$ waveforms are not as accurate as the EOB$_{\rm HH}$
waveforms, which include the subdominant modes. This is true even for
comparable-mass systems, where the subdominant modes only minimally
contribute to the overall waveform power.  The bias introduced by
neglecting additional harmonics is not due to missing waveform power
as much as it is caused by phase differences between a quadrupole-only
template and the full NR data, as coherent matched-filtering analyses
are typically more sensitive to phase than amplitude.

The parameter estimation accuracy of the EOB$_{\rm HH}$ model up to
$\rm{SNR}\sim50$ exceeds expectations from Ref.~\cite{Pan:2011gk}, as
can be seen by focusing on rows 2 and 7 in Table~\ref{tbl:intrinsic}. 
Here we find systems chosen specifically to compare with Fig.~15 
in Ref.~\cite{Pan:2011gk} where, based on the accuracy requirement proposed in
Refs.~\cite{Lindblom:2009ux,Lindblom:2010mh}, they predicted that
systematic error could exceed statistical error at {\it
  single-detector} SNR $\sim 35$ ($q=1$, $M=50\MSun$) and $\sim 11$
($q=6$, $M=56\MSun$).  \footnote{The accuracy criterion used in
    Refs.~\cite{Pan:2011gk,Lindblom:2009ux,Lindblom:2010mh} is a ``sufficient'' but ``not
    necessary'' requirement for parameter estimation, and it does not say 
   which of the binary parameters will be biased and how large the bias will be. 
   Thus, the authors of Ref.~\cite{Pan:2011gk} 
    were making conservative judgments about the waveform accuracy.}

Our analysis uses the LIGO/Virgo network of detectors, as opposed to the 
single-detector studies from Ref.~\cite{Pan:2011gk}. This difference will not heavily impact the 
results, as it is the measurement of \emph{intrinsic} parameters that is most affected 
by differences in the waveform model due to both the accuracy with which they are measured, 
and the way they are encoded in the phase evolution of the signal.  
Measurement of the intrinsic parameters is not greatly influenced by the inclusion of 
additional detectors in the network (at a fixed SNR). 
Our findings show that, even at SNRs that are rather high for an
expected LIGO detection, the EOB$_{\rm HH}$ model introduces
systematic errors that differ by $\lesssim 1\sigma$ from the injected
parameters.   

\begin{table*}[htdp]
\begin{center}
\begin{tabular}{lccc|cc|cccc|cccc|cccc}
\hline\hline
&&&&&& \multicolumn{4}{c|}{$p(\rm{EOB}|\rm{EOB})$} &  \multicolumn{4}{c|}{$p(\rm{EOB_{22}}|\rm{NR})$} & \multicolumn{4}{c}{$p(\rm{EOB_{HH}}|\rm{NR})$} \\ 
$q$ & $M$ $(\MSun)$ & $f_{\rm low}$  (Hz)& SNR & $\sigma_{\sin \delta}$ & $\sigma_{\alpha}$ (rad) & $\sigma_{\ln D_L}$  & $\delta\beta_{\ln D_L}$ & $\delta\beta_{\sin \delta}$ & $\delta\beta_{\alpha}$ & $\sigma_{\ln D_L}$ & $\delta\beta_{\ln D_L}$ & $\delta\beta_{\sin \delta}$ & $\delta\beta_{\alpha}$ & $\sigma_{\ln D_L}$  & $\delta\beta_{\ln D_L}$ & $\delta\beta_{\sin \delta}$ & $\delta\beta_{\alpha}$ \\
\hline
1 & 50& 30 & 12 &  0.06 & 0.04 & --- & --- & --- & --- & --- &  --- & --- & --- & --- & --- & --- & ---\\
1 & 50& 30 & 48 & 0.03 & 0.01 & 0.10 & 0.11 & 0.01  & 0.09 & 0.12 & 0.12 & 0.17 & 0.03 & 0.20 & 0.64 & 0.07 & 0.30\\

2$^*$ & 23& 60 & 12 & 0.06 & 0.03 & 0.24 &  0.02 & 0.04 & 0.02 & 0.27 & 0.28 & 0.02 &0.06 & 0.25 & 0.30 & 0.05 & 0.08\\

2 & 51& 30 & 12 & 0.06 & 0.04 & 0.28 & 0.03 & 0.08 & 0.06 & 0.29 & 0.21 & 0.28 &0.07& 0.27 & 0.60 & 0.15 & 0.33\\

2 & 51& 30 & 48 & 0.03 & 0.02 & 0.12  & 0.10 & 0.10 & 0.08 & 0.14 & 0.11 & 0.19 &0.11& 0.11 &  0.07 & 0.04 & 0.05\\

6 & 56& 30 & 12 & 0.07 & 0.05 & 0.29  & 0.07 & 0.04 & 0.21 & 0.29 & 0.28 & 0.14 &0.09& 0.25 &  0.40 & 0.41 & 0.39\\

6 & 56& 30 & 48 & 0.03 & 0.02 & 0.14  & 0.29 & 0.10 & 0.07 & 0.19 &  0.28 & 0.14 &0.09& 0.09 & 0.78 & 0.27 & 0.21\\

6$^*$ & 120& 10  & 12 & 0.08 & 0.05 & 0.21  & 0.23 & 0.28 & 0.13 & 0.30 & 0.86 & 0.37 &0.54& 0.22 & 0.36 & 0.02 & 0.16\\
\hline
\end{tabular}
\caption{{\small Same as Table~\ref{tbl:intrinsic}, except here we show a subset of the extrinsic parameters corresponding to the binary's location.  Because of their similarity between each run, the statistical errors are displayed once but apply to each example. 
Results for the $q=1$, ${\rm SNR}=12$ example are omitted due to the failure of our $\delta\beta$ statistic.  
}}
\label{tbl:extrinsic}
\end{center}
\end{table*}
The extrinsic parameters, on the other hand, are inferred mostly from
the overall amplitude of the waveform, which is not as well-measured
as phase, and the time of arrival of the signal at each detector. 
We thus expect that the extrinsic parameters, determined with 
lower fidelity than the masses, will be better able to tolerate
small differences between template waveform models within the
statistical error.  Adding additional detectors to the network 
dramatically improves the statistical error for extrinsic parameters,
mostly due to the increased baseline~\cite{Veitch:2012df}, but not to
the point of becoming influenced by the waveform systematics.

Indeed, we find that the relative systematic biases for extrinsic parameters 
are generally smaller that those of the intrinsic parameters. For systems with $q \geq 2$, regardless
of the SNR or the EOB model, the systematic errors are consistently
smaller than the statistical errors, even when the $(2,2)$-only 
waveform is used as the template.  This is evident in
Fig.~\ref{fig:posteriors}, where the $D_L$, $\sin\delta$, and
$\alpha$ posteriors are nearly indistinguishable, despite the
significant difference in the residual left behind by the waveform
model, as shown in the bottom-right panel containing the
logarithmic likelihood distributions.  The same can not be said for the equal-mass
cases (top two rows in Table~\ref{tbl:extrinsic}), where we
encounter a subtle effect from our choice of statistic,
$\delta\beta$. 

\begin{figure*}[htbp]
   \centering
   \includegraphics[width = \linewidth]{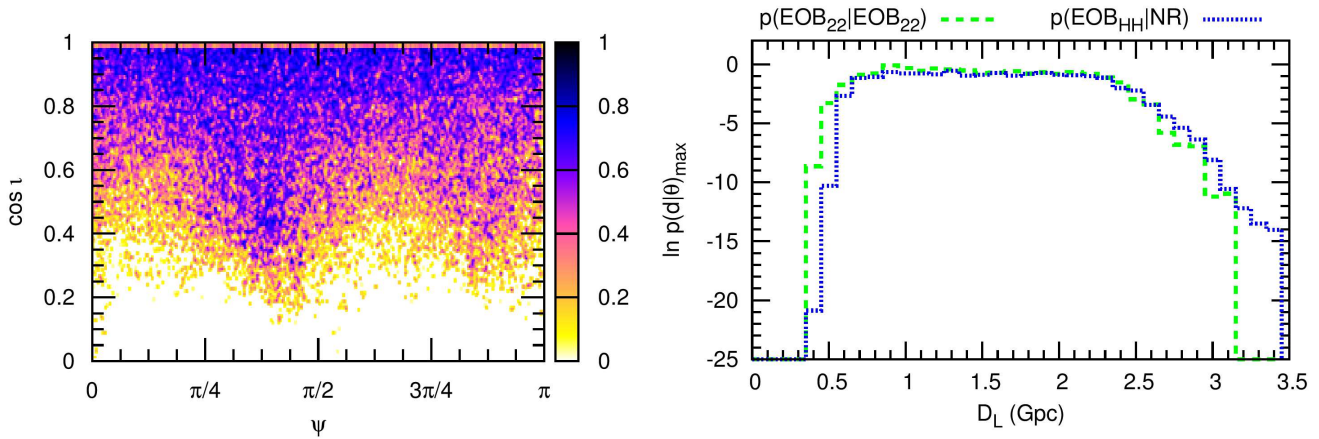} 
   \caption{{\small Selected results from the $q=1$, ${\rm SNR}=12$ run to exhibit the breakdown of
   $\delta\beta$ as a useful statistic.  The left-hand panel shows the 2D marginalized posterior
   for the orientation angles $\psi$ and $\cos\iota$, with darker colors corresponding to higher
   probability density.
   The right-hand panel displays the maximum logarithmic likelihood as a function of $D_L$, which is effectively
   uniform between $\sim 0.5$ and $\sim 2.25$ Gpc.}}
    \label{fig:orientation}
\end{figure*}

For the SNR = 12, equal-mass case, the $\delta\beta$ statistic breaks
down and results are omitted from the table.  At such a low signal
strength, the orientation parameters are very poorly measured, with
the polarization angle $\psi$ effectively unconstrained.  These large
measurement uncertainties cascade through the 1D posteriors via strong
$\psi-\iota$ and $\iota-D_L$ covariances.  We are left with an
unconstrained $D_L$ distribution that is poorly characterized by the
variance, and large stochastic variation from one Markov chain run to
the next as to where the MAP parameters lie.  This degeneracy is evident in
Fig.~\ref{fig:orientation}, where we show the 2D marginalized posterior
distribution function of the $\psi - \cos\iota$ plane (left panel)
from a $p({\rm EOB}_{\rm HH}|{\rm NR})$ run, and the maximum
logarithmic likelihood found in the Markov chain for different bins in $D_L$
space (right panel) from both $p({\rm EOB}_{\rm 22}|{\rm EOB}_{\rm
  22})$ and $p({\rm EOB}_{\rm HH}|{\rm NR})$.  We see a virtually flat
distribution of the maximum logarithmic-likelihood values between $\sim 0.5$
to $\sim 2.25$ Gpc, with well over half of the allowed parameter space
in the $\psi - \cos\iota$ plane receiving significant posterior
support.  The injected value of $D_L$ was near 1 Gpc.

The over-density at $\{\psi,\cos\iota\} \simeq \{\pi/3, 0.5\}$ in the left panel corresponds
to the injected parameter values, with the $\pi/2-{\rm shift}$
degenerate mode still appearing despite the inclusion of subdominant 
modes.  Recall that this is an equal-mass system, where the
subdominant modes are the least noticeable.  The over-density at
$\cos\iota \sim 1$ is due to the Markov chain preferring template
waveforms with minimal contribution from subdominant modes (to
match the strictly equal-mass injection) as the sampler explores
higher mass-ratios, up to $q\sim3$ in this case. Systems with
$\cos\iota \equiv \hat{\mathbf{k}}\cdot \hat{\mathbf{L}} = 1$ are
face-on and it is this configuration where the subdominant modes are 
least prominent.

\subsection{Dependence of the results on the inclination angle}

Due to the high computational cost of each MCMC run, we are not able to
Monte Carlo over a large population of binary systems.  We instead
have chosen extrinsic parameters away from the extremes of
parameter space.  This means sky locations that are away from nulls in
any detector's response, and inclinations ($\iota = \pi/3$) that
were not edge- or face-on with respect to the observer's line of
sight.

One of the more interesting results from this study is the impact of
the subdominant modes on the parameter-estimation capabilities of
ground-based detectors.  The role that the additional modes play in
the waveform depends heavily on both the mass-ratio and the
orientation of the binary --- edge-on systems have the largest
contribution from the additional modes, while face-on systems are most
dominated by the $(2,2)$ mode. It is therefore possible that, for
some more extreme orientations, systematic biases could become large
because of the increased importance of the additional modes. 

To allay this concern we performed a series of MCMC runs on a system
where the subdominant modes would play an important role, exploring
the edges of orientation space for each run.  We chose the $M=120\MSun$,
$q=6$ system (row 8 in Tables~\ref{tbl:intrinsic} and
\ref{tbl:extrinsic}) and analyzed three different orientations:
edge-on ($\iota = \pi/2$), face-on ($\iota = 0$), and moderate tilt
($\iota = \pi/3$).  We compare the $\Delta \ln M$ and $\Delta \ln
\mathcal{M}$ posteriors for each of these systems using the EOB$_{\rm
  HH}$ model as a template to study data containing an NR waveform
injected at SNR$=12$.  
\begin{table}[htdp]
\begin{center}
\begin{tabular}{c|cc|cc|c}
\hline\hline
$\iota$ & $\sigma_{\ln M}$ & $\delta\beta_{\ln M}$ & $\sigma_{\ln \mathcal{M}}$ & $\delta\beta_{\ln \mathcal{M}}$ & \%HH\\
\hline
$0$ & 0.03 & 0.10 & 0.04 & 0.12 & $\sim$0 \\
$\pi/3$ & 0.02 & 0.29 & 0.03 & 0.06 & 7.5 \\
$\pi/2$ & 0.02 & 0.18 & 0.03 & 0.01 & 10 \\
\end{tabular}
\caption{{\small Fractional systematic errors and statistical errors
    for $\ln M$ and $\ln \mathcal{M}$ when $M=120\MSun$, $q=6$,
    ${\rm SNR}=12$ and for three different inclinations: Edge-on
    ($\iota = \pi/2$), face-on ($\iota = 0$), and an intermediate
    orientation ($\iota = \pi/3$).  We also include the percentage that
    the additional modes (HH) contribute to the total
    SNR.}}
\label{tbl:inclination}
\end{center}
\end{table}
The results in Table~\ref{tbl:inclination} show
the fractional systematic error well below unity for each orientation
regardless of the inclination angle.  We also include the percentage
of the total SNR that comes from the subdominant harmonics (HH).  
This result confirms
that the parameter-estimation accuracy of the EOB model is robust to
different orientations, and thus different strengths of the additional
modes.

\subsection{Simulating a detection}

All of the above results have been performed on simulated data
that do not contain any noise, but do include the noise PSD in the
inner product defined in Eq.~(\ref{eq:nwip}).  Thus the posteriors that
we generate are not representative of a probability density function
for an actual GW measurement, but instead are the hypothetical
averaged measurements of the same system in an ensemble of noise
realizations~\cite{Nissanke:2009kt, Cornish:2011ys,
  Vallisneri:2011ts}.  To more realistically
demonstrate the parameter-estimation capabilities of advanced
ground-based interferometers, we want now to 
simulate a single LIGO/Virgo detection. 

To that end, we use again the binary configuration with $q=2$, $M = 51\MSun$  
and network SNR of 12, but now add stationary, gaussian noise to the NR
waveform using the same PSD as in the noise-free study.  The
resultant posteriors are then representative parameter-estimation
products, subject to the following important caveats:
\begin{itemize}
\item We use the same PSD for each detector when, in practice, each interferometer 
  will have different sensitivity at any given time.  Furthermore, the
  Virgo design sensitivity is not identical to LIGO (although it is
  qualitatively similar).  We also effectively introduce a noise
  ``wall'' at 30 Hz to account for the limited duration of the NR
  data.
\item We do not include any calibration errors in the waveform
  injections, which could prove to be a significant contribution to the
  overall parameter-estimation error budget~\cite{Vitale:2011wu}.
  Furthermore, we do not account for intrinsic error in the 
  NR waveforms.
\item We recognize that simulated additive gaussian noise is different from 
injecting waveforms into real LIGO/Virgo noise~\cite{Raymond:2009cv}.
\end{itemize}

\begin{figure*}[htbp]
   \centering
   \includegraphics[width = 0.8\textwidth]{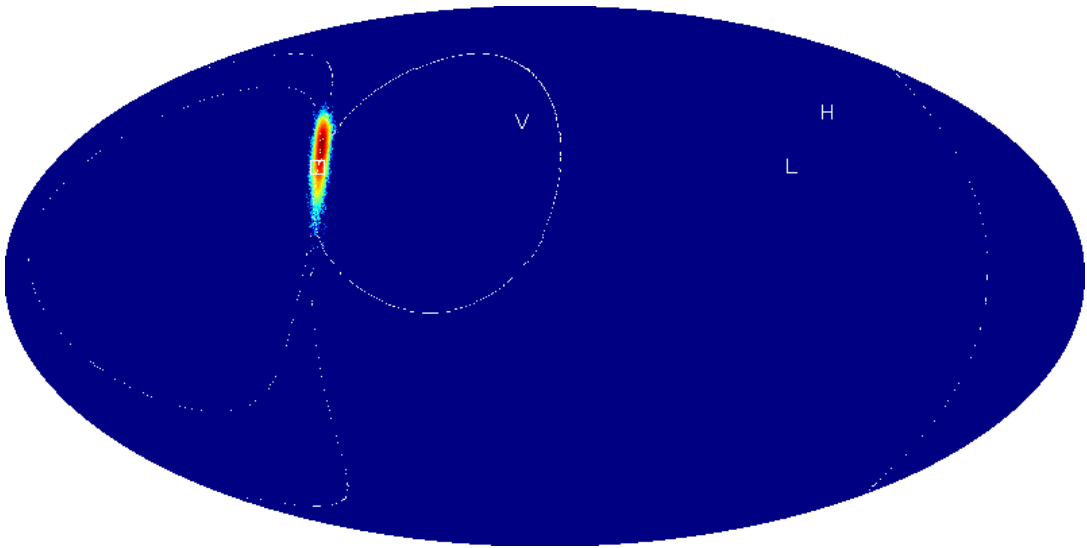} 
   \caption{{\small 2D marginalized posterior distribution function for the sky location of a $q=2$, $M = 50\MSun$, binary with ${\rm SNR} = 12$ injected into simulated stationary gaussian noise colored by the Advanced LIGO noise power spectral density.  The red, yellow, and cyan regions roughly correspond to 1, 2, and 3$\sigma$ confidence regions.  The white box represents the injected sky location of this source.  The position of each interferometer in the network is projected onto the sky, labeled with H (LIGO Hanford), L (LIGO Livingston), and V (Virgo).  The white, dashed lines show the locations that yield the same time-delay between each pair of detectors for the injected sky position. 
   }}
   \label{fig:2Dsky}
\end{figure*}

\begin{figure}[htbp]
   \centering
   \includegraphics[width = .5\textwidth]{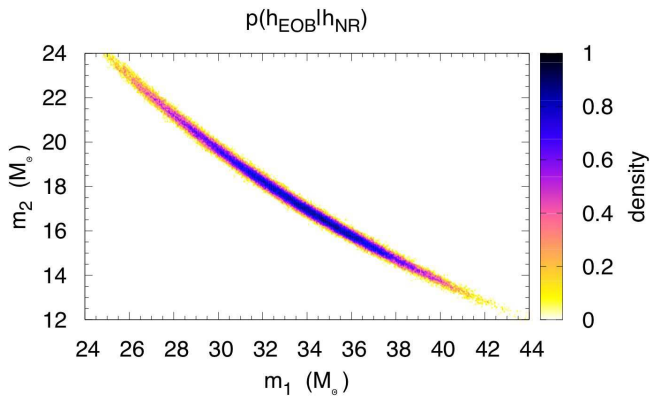} 
   \caption{{\small 2D marginalized posterior distribution function for the individual masses $m_1$ and $m_2$ of a $q=2$, $M = 51\MSun$, binary with ${\rm SNR} = 12$,
   injected into simulated stationary gaussian noise colored by the Advanced LIGO noise power spectral density.  The injected values for $\{m_1,m_2\}$ were $\{34,17\}\MSun$.  
   }}
   \label{fig:m1m2}
\end{figure}

\begin{figure}[htbp]
   \centering
   \includegraphics[width = .5\textwidth]{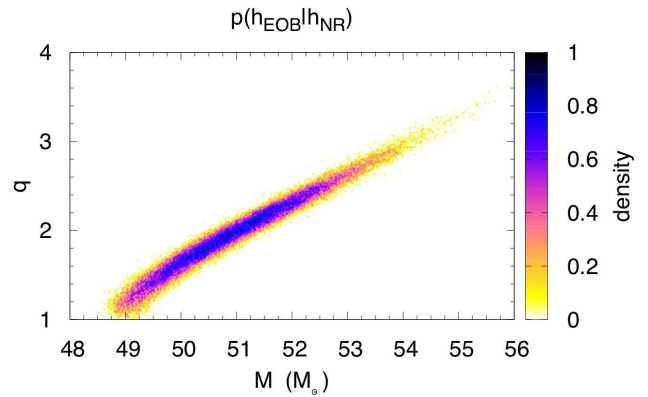} 
   \caption{{\small Same as Fig.~\ref{fig:m1m2}, but now depicting total mass $M$ and mass-ratio $q$. }}
   \label{fig:Mq}
\end{figure}

For this study we find the 2D marginalized posterior distribution
functions to be of the most interest.  We show results for the
sky location in Fig.~\ref{fig:2Dsky} and mass parameters in 
Figs.~\ref{fig:m1m2} and~\ref{fig:Mq}.

In Fig.~\ref{fig:2Dsky}, the sky-location posterior is shown in a Molweide projection with the
detector locations projected on to the celestial sphere. The white,
dotted lines show the circles of constant time delay between each pair
of detectors.  The posterior should sit at intersections of these lines, and the
principal axis should lie along a line.  A small white square is
included, centered on the injected position. The injected values for
the sky location are contained within the $\sim 63\%$ confidence
interval of the posterior (the red region of the error ellipse).  The
injected sky location was chosen to be a region where the SNR in each
detector was roughly equivalent.

Of more pertinence to this study are the mass posteriors.  While $\ln
M$ and $\ln \mathcal{M}$ are the most convenient parameters for the
MCMC sampler, being the most orthogonal, they are not of
the most interest to the wider astrophysical community.  A better data
product would be posteriors on either the individual masses $m_1$ and
$m_2$, or the total mass $M$ and mass-ratio $q$.  In post-processing we take
the MCMC chains and compute the relevant mass parameters at each step
in the chain.  The injected component black holes have masses $m_1 = 34\MSun$ 
and $m_2 = 17\MSun$.  We show the 2D marginalized posterior distribution
functions for the the $m_1\mbox{--}m_2$ (Fig.~\ref{fig:m1m2}), 
and $M\mbox{--}q$ plane (Fig.~\ref{fig:Mq}), where the color
corresponds to the posterior
density.  

These figures give a good depiction of just how correlated the mass
parameters are with one another, and how much of parameter space is
supported by the chain in part due to that strong correlation.  For
example, the $M\mbox{--}q$ plane has significant support for mass-ratios
between 1 and 3, compatible with previous LIGO MCMC studies 
using PN waveforms (e.g.,~ see Ref.~\cite{vanderSluys:2007st}). 
These are the type of parameter-estimation 
products that the astrophysics community can anticipate as the advanced 
detectors come on line in the coming years.
 
\section{Results for space-based detectors}
\label{sec:LISA}

For EOB waveforms that include subdominant modes, we have found relatively small systematic errors in 
parameter-estimation results for ground-based observations with $\rm{SNR}<50$. Because ground-based 
GW instrument rates are limited by sensitivity, higher-SNR events are exceedingly unlikely in the first generation of detections. 
Proposed space-based instruments will be
sensitive to supermassive black-hole (SMBH) mergers out to cosmological scales, such that a significant fraction of
detected events may have $\rm{SNR}>100$. Space-based instruments are typically sensitive to these events over a broad bandwidth
covering a large number of cycles leading up to merger~\cite{LISASciCase:2009,eLISA}.  Such observations will make
much greater demands on the accuracy and efficiency of inspiral-merger-ringdown waveform templates.  Though considerable
effort has gone into estimating the ability of space-based instruments to measure astrophysical parameters assuming accurate 
waveforms, very little has been done to assess the template requirements for these future observations. Here we
make a limited exploration of 
this capability with current numerical-relativity and EOB waveforms.

Of several proposed space-based GW interferometer instruments \cite{LISA,Anderson:2012sgo,eLISA,Kawamura:2008zza}, 
the best-studied concept is the classic
LISA mission~\cite{LISA} . While acknowledging that there is currently considerable 
uncertainty about when and how the first space-based GW instrument 
will be developed, we choose to study the classic LISA configuration to make 
contact with the large body of work that
has already been dedicated to black-hole merger parameter
estimation (e.g., see Refs.~\cite{Berti:2007zu,Thorpe:2008wh,McWilliams:2009bg, McWilliams:2010eq, McWilliams:2011zs}). To compare with other concepts, the most relevant
alteration from the classic LISA design is the
arm length (e.g., from $5$ Gm for classic LISA down to $1$ Gm for eLISA), which sets
the overall scale for parameter-estimation capabilities; our 
results for a more modest detector configuration would be very
similar to those for classic LISA, after appropriately rescaling the total mass of the 
black-hole system.

We follow here the same procedure outlined for the
LIGO/Virgo studies in Sec.~\ref{sec:LIGO}, where NR waveforms are injected
into simulated noise-free data, and the signals are analyzed using the
EOB model as a template.  Because we saw significant bias in the
EOB$_{22}$ model at SNR $\sim 50$ it is safe to assume that
those errors will only grow with SNR, and so we focus these runs only
on the EOB$_{\rm HH}$ model.  

The duration of the available NR data restricts us to brief LISA observations, for which we can apply the
static limit for the detector; thus we neglect LISA's orbital motion during
the observation time.  Consistent with this, we focus on systems of mass $3\times10^7 \MSun$ at the high end of
LISA's sensitive range. For such observations the maximum frequency attained by the merger signal
is well below the transfer frequency of the detector (when the
wavelength of the GW signal is comparable to the size of the
detector).  In this low-frequency, static regime, the instrument
response is equivalent to two 60$^\circ$ Michelson interferometers,
co-located, and misaligned by $\pi/3$ radians.  The antenna patterns 
for this configuration, and the discussion of the two
limits applied here, can be found in Ref.~\cite{Cornish:2002rt}.

\begin{figure*}
   \centering
   \includegraphics[width = \linewidth]{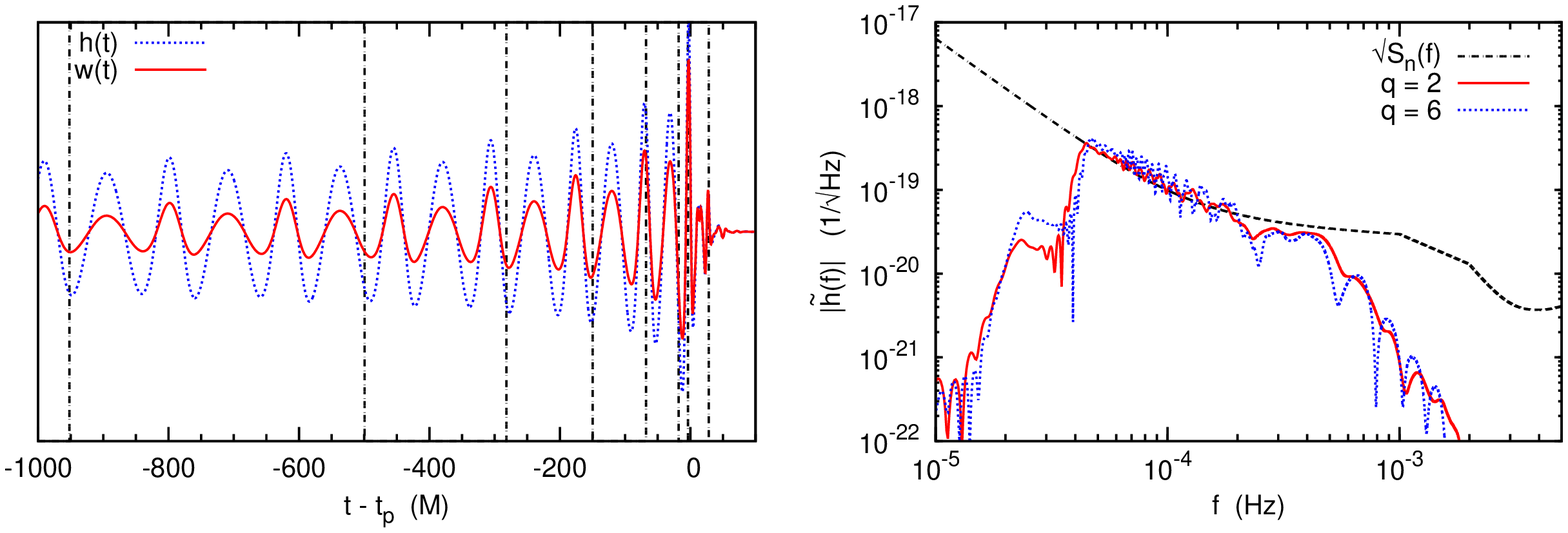} 
     \caption{\small{ Same as Fig.~\ref{fig:ligo_psd} now showing the LISA examples using $M = 3\times10^7\MSun$ at $\rm{SNR}=100$.  The time-domain waveform in the left-hand panel is for the $q=6$ case.
   }}
   \label{fig:lisa_psd}
\end{figure*}
We consider mass-ratios in the range $2\leq q\leq6$, observed at SNR=100, which would make these unusually distant 
for LISA observations. 
As shown in Fig.~\ref{fig:lisa_psd}, the power-spectral density is similar over this mass-ratio range
with more structure at $q=6$.  The noise-weighted waveforms for these cases are most comparable to the
largest-mass $120 \MSun$  LIGO case that we studied.  Space-based instruments are not expected to have
a strong power-law slope like the seismic noise wall in the LIGO sensitivity curve, meaning that even
for large masses there is a softer degradation of sensitivity going back to the early portions of the
signal, making our $f_\mathrm{min}$ cut somewhat more artificial here.

\begin{table*}[htdp]
\begin{center}
\begin{tabular}{cccc|ccccc|ccccc}
\hline\hline
&&&& \multicolumn{5}{c|}{$p(\rm{EOB_{HH}}|\rm{EOB_{HH}})$} & \multicolumn{5}{c}{$p(\rm{EOB_{HH}}|\rm{NR})$} \\

$q$ & $M\ (\MSun)$ & $f_{\rm low}\ ({\rm Hz})$ & SNR & $\sigma_{\ln M}$ & $\sigma_{\ln \mathcal{M}}$ & $\delta\beta_{\ln M}$ & $\delta\beta_{\ln \mathcal{M}}$ & ${\rm SNR}_{\rm res}$ & $\sigma_{\ln M}$ & $\sigma_{\ln \mathcal{M}}$ &  $\delta\beta_{\ln M}$ & $\delta\beta_{\ln \mathcal{M}}$  & ${\rm SNR}_{\rm res}$\\
\hline

2 & $\   3\times 10^7$ & $5\times 10^{-5}$ & 100 & $2\times10^{-3}$ & $2\times10^{-3}$ & 0.09 & 0.08 & 0.5 & $2\times10^{-3}$ & $2\times10^{-3}$ & 1.19 & 1.27 & 6.5\\

3 & $\   3\times 10^7$ & $5\times 10^{-5}$ & 100 & $2\times10^{-3}$ & $3\times10^{-3}$ & 0.02 & 0.04 & 0.7 & $3\times10^{-3}$ & $2\times10^{-3}$ & 0.59 & 2.82 & 7.7\\

4 & $\   3\times 10^7$ & $5\times 10^{-5}$ & 100 & $3\times10^{-3}$ & $2\times10^{-3}$ & 0.10 & 0.09 & 0.5 & $3\times10^{-3}$ & $3\times10^{-3}$ & 1.31 & 3.02 & 9.6\\

6 & $\ 3\times 10^7$ & $5\times 10^{-5}$ & 100 & $2\times10^{-3}$ & $2\times10^{-3}$ & 0.16 & 0.27 & 0.5 & $3\times10^{-3}$ & $2\times10^{-3}$ & 0.62 & 0.92 & 11.5\\

\hline
\end{tabular}
\caption{{\small Same as Table~\ref{tbl:intrinsic} but for SMBH mergers as seen by LISA.
Here we also include the residual SNR after the MAP waveform is regressed from the data. 
}}
\label{tbl:LISA}
\end{center}
\end{table*}
Generally the higher SNRs of our nominal LISA observations would predict larger bias from systematic 
errors in the template waveforms. Because of the differences between the sensitivity curves and 
response for LISA and LIGO, however, it is not straightforward to scale up such expectations.  
Table~\ref{tbl:LISA} shows our results for the parameter biases for mass $M$ and chirp mass $\mathcal{M}$.  Unlike
the LIGO results, the biases here are already statistically significant in most cases for LISA observations at SNR=100,
reaching a few times the statistical error level.  
In Table~\ref{tbl:LISA} we also provide the SNR of the residuals after the MAP waveforms are removed from 
the data.  These residuals with SNR $>6$ would be detectable and could therefore lead to biases in 
estimates of overlapping signals. Residuals at this level would also limit the utility of the current waveform templates
for studies aimed at testing general relativity~\cite{Li:2011cg,Vallisneri:2012qq}.

For LISA, however, such a system would have to
be exceedingly distant, at redshift $z>20$ or more, in order to expect an SNR as small as 
100~\cite{McWilliams:2009bg}.  
Actual SNRs could be as much as 100 times larger, and we would expect correspondingly larger relative biases and 
residuals. Full interpretation of LISA data would thus require higher levels of template accuracy. 
Even the level of errors in the numerical simulations used here 
in place of the exact predictions of general relativity 
are far too large to avoid biasing such high-SNR measurements. 

That said, while the statistical error should decrease
  linearly with increasing SNR (causing $\delta\beta$ to grow), the
  systematic error should remain approximately the same, and the
  absolute biases in the mass parameters are small.  Reading from
Table~\ref{tbl:LISA}, we see that the systematic errors in $M$ and
  $\mathcal{M}$ are $\lesssim1\%$.  Converting these into the
  individual masses, the biases on $m_1$ and $m_2$ are at most a few
  percent.  While the MAP mass parameters may end up many $\sigma$
  away from the true values and would thus fall short of an
    optimal analysis of LISA data, such $\sim1\%$ errors in the
  masses are still small in astrophysical terms and may 
  have little impact on key inferences made about the
  supermassive black-hole population.  For example, when trying to
  constrain black-hole formation scenarios using simulated eLISA BBH catalogs,
Amaro-Seoane \emph{et al.}~\cite{eLISA} (based on the procedure in
  Ref.~\cite{Sesana:2010wy}) endured statistical errors as high as
  $\sim1\%$ and were still able to easily discriminate between black
  hole seed models.  

The limitations of our current analysis prevent us from providing any detailed indication of how far
template accuracy must be improved for space-based observations.
For the detector configuration used
here, the \emph{extrinsic} parameter estimation at SNR $=100$ is actually worse than
that for the ground-based detectors.  Space-based detectors rely on
the long duration of the signals, the amplitude and Doppler
modulations caused by the orbital motion, and finite-arm-length
effects, to break degeneracies among the system parameters.  We are
limited here to working in a regime where all of those features are missing
from our instrument model, so our classic LISA response is more like a
two-detector co-located LIGO network operating at significantly lower
frequencies than the existing ground-based detectors. 

More complete analysis will require either dramatically longer-duration numerical simulations
or a suitably well-controlled way of matching analytical and NR waveforms at 
low frequency that does not add as much systematic error as the template models being tested.  
We would also require more computationally efficient signal generation to successfully study 
template biases with long-duration waveforms.

\section{Conclusions and discussion}
\label{sec:conclusions}

In this paper, we have produced the first measurement of systematic
errors introduced by using EOB templates to analyze NR waveforms. 
Our study's main focus was on stellar-mass BBHs observed by Advanced LIGO and Virgo.
We also considered supermassive black-hole mergers detectable by 
space-based interferometers like LISA. 
We have injected NR waveforms into simulated data and have used an MCMC sampler to
characterize the posterior distribution function for the astrophysical
parameters.  Our metric for assessing the size of the systematic error
is to compare the offset between the injected and best-fit parameters
to the statistical error characterized by the standard deviation of
the 1D marginalized posteriors.

For the stellar-mass systems we have investigated, when including the
subdominant modes, we find systematic biases consistently comparable
to, or smaller than, the statistical errors for mass-ratios up to
$q=6$ and SNRs $\lesssim 50$.  We have tested these waveforms in the
most stringent way possible, simulating high-SNR events where the
merger (the least reliable part of the waveform calculations) is
peaking in the most sensitive band of the detector, and the
higher-frequency modes contribute significantly to the overall signal
power. For the $q=6$ waveforms weighted by the detector PSD, the fraction of power contained in the
subdominant modes is 11$\%$ and 16$\%$ for the $M=56 \MSun$ and $M=120
\MSun$ systems, respectively. 
We also tested low-mass systems
($M\sim20\ \MSun$) to better represent likely Advanced LIGO/Virgo
detections. 

In all of these examples, the bias introduced by the EOB
waveform in Ref.~\cite{Pan:2011gk} was \emph{at worst} comparable to the statistical
errors.
For several of our examples chosen specifically to compare with
that paper, we find that 
the EOB waveforms accurately recover binary parameters 
at SNRs higher than were predicted there using 
the (deliberately) conservative accuracy requirements of 
Refs.~\cite{Lindblom:2009ux,Lindblom:2010mh}.

Matched-filtering analyses are most sensitive to the phase of the
signal, and it is the phase of the waveform that is most influenced by
different choices of the model.  It is therefore predictable that the largest biases appear in the mass parameters.  On the other hand,
the extrinsic parameters have comparatively less impact on the shape
of the signal --- the distance comes in as an overall amplitude
scaling, and the sky location is (for ground-based interferometers)
predominantly determined through triangulation based on time
delays between detectors.  Therefore a model waveform used for
parameter estimation has much more room for error if, for instance,
the location of the binary is the primary interest (say, for optical
counterpart searches) and the requirements on the phase-matching are
not as severe.  

While the results here are undoubtedly positive, there is still work
to be done in waveform modeling.  We are only testing the EOB 
waveforms over the last $30\mbox{--}40$ GW cycles before merger and
there is no guarantee that longer waveforms will not accumulate larger
phase errors during the early portion of the inspiral.  Furthermore,
this study neglects significant aspects of the waveform structure
related to black-hole spins and orbital eccentricity.  A similar study will need to be
performed with long-duration, spinning systems once both the NR and
EOB waveforms are prepared for that test. 
It will also be valuable to test the EOB waveforms over a broader class of NR simulations, 
including those that were not used to calibrate the template model.  
Moreover, we have assumed 
in this paper that the NR waveforms were exact but, as discussed 
in Sec.~\ref{sec:waveforms}, this is not the case. One possibility for taking 
the numerical error into account is to inject NR waveforms computed at different resolutions 
and/or extracted at different radii and measure the EOB systematic biases in 
each case. The difference between these biases can provide us with an estimate 
of the intrinsic error caused by the NR waveforms deviating from the exact solution 
in general relativity. Finally, we do not consider the effects of real detector noise or calibration errors in the data, both
of which could prove to be a significant contribution to the overall
error budget~\cite{Raymond:2009cv,Vitale:2011wu}.

While EOB waveforms that include subdominant modes were found to have relatively small systematic errors in 
parameter-estimation results for ground-based observations with SNR $<50$, proposed space-based instruments are
sensitive to SMBH mergers with SNR $>100$.  
For these scenarios, we find statistically significant biases in the mass parameters for mass-ratios in the range $2\leq q\leq6$ observed at $\rm{SNR}=100$, on the order of $\sim1\%$ for the component masses of the system. 
However, as discussed in Sec.~\ref{sec:LISA}, systematic errors introduced by
the EOB templates are small enough to still place strong constraints on the
population of supermassive black holes in the Universe.

In the LISA examples, the residual power SNR is $>6$,
sufficient to compromise  
both parameter estimation of overlapping signals 
and studies aimed at testing general relativity.  
Space-based GW data analysis will require 
more accurate templates, grounded in even more accurate numerical simulations. 

A more complete analysis is needed, but will require either dramatically longer-duration numerical simulations
or a suitably well-controlled way of matching analytical and NR waveforms at low frequency.  
We would also require more computationally efficient signal generation to successfully study 
template biases with long-duration waveforms.

\section{Acknowledgments}

We would like to thank Ryan Lang, Sean McWilliams, Yi Pan, and Ira Thorpe 
for very useful discussions. We also thank the Caltech-Cornell-CITA 
collaboration for providing us with the NR waveforms used in this work. 

The authors acknowledge support from the NASA Grants 08-ATFP08-0126, 11-ATP11-0046,
J.B and B.K. also acknowledge support from 09-ATP09-0136. 
A.B. also acknowledges support from the NSF Grants No. PHY-0903631 and 
PHY-1208881.  

The MCMC runs were carried out using resources from the NASA Center 
for Climate Simulation at Goddard Space Flight Center, and the Nemo cluster 
supported by National Science Foundation awards PHY-0923409 and PHY-0600953 to
UW-Milwaukee. 

\bibliography{papers}

\end{document}